**The spatial scan statistic: A new method for spatial aggregation of categorical raster maps**

J.W. Coulston†, N. Zaccarelli‡, K.H. Riitters§, F.H. Koch† and G. Zurlini‡*

† North Carolina State University, Department of Forestry and Environmental Resources
Campus Box 8008, Raleigh, NC, 27695-8008, USA
‡ University of Salento, Department of Biological and Environmental Sciences and Technologies, 73100 Lecce, Italy
§ USDA Forest Service, Southern Research Station, Forestry Sciences Laboratory
3041 Cornwallis Road, Research Triangle Park, NC, 27709 USA

* To whom correspondence should be addressed, giovanni.zurlini@unisalento.it

**Abstract**

Multiple-scale and broad-scale assessments often require rescaling the original data to a consistent grain size for analysis. Rescaling categorical raster data by spatial aggregation is common in large area ecological assessments. However, distortion and loss of information are associated with aggregation. Using a majority rule generally results in dominant classes becoming more pronounced and rare classes becoming less pronounced. Using nearest neighbor techniques generally maintains the global proportion of each category in the original map but can lead to disaggregation. In this paper we implement the spatial scan statistic for spatial aggregation of categorical raster maps and describe the behavior of the technique at the local level (aggregation unit) and global level (map). We also contrast the spatial scan statistic technique with the majority rule and nearest neighbor approaches. In general, the scan statistic technique behaved inverse the majority rule approach in that rare classes rather than abundant classes were preserved. We suggest the scan statistic techniques should be used for spatial aggregation of categorical maps when preserving heterogeneity and information from rare classes are important goals of the study or assessment.

**1 Introduction**

In order to quantify spatial heterogeneity satisfactorily and detect characteristic scales of landscapes, it is widely recognized that landscapes should be examined at multiple scales. The role played by scale in landscape analysis can be described by distinguishing between the scale of observation (the scale at which the natural world is translated into data) and the scale of analysis (the scale at which patterns are revealed from the data) (Li and Reynolds 1995). Often they are not identical. The scale of observation stems directly from the characteristics of the system studied, the questions asked, and the data collection protocols. The scale of observation is an inherent property of the data and is often changed into scale of analysis because of the difficulties in collecting data at multiple scales of observation. The scale of analysis is determined by the interaction of the original scale of observation with the methods used for data transformation, including aggregation, magnification, and resampling.

Large-area (regional, national extent) ecological assessments rely on a variety of spatial information across a range of scales (grain sizes). For instance, land cover or land use data commonly serve as the basis for such assessments. Many of these spatial data sets are derived from satellite imagery collected by sensors with different grain sizes, such as the Landsat Thematic Mapper (30m), the Moderate Resolution Imaging Spectroradiometer (250m to 1000m), and the Advanced Very High Resolution Radiometer (1000m). Other frequently used raster data such as digital elevation models, interpolated climate data, and soil data are also available in varied grain sizes. Combining these data for national or regional ecological assessments often requires re-scaling or aggregating to the same spatial resolution to enable efficient computation and simplify the interpretation of analytical outputs (.e.g., Coulston and Riitters 2005, Verburg and Veldkamp 2004). Unfortunately, distortion and the loss of potentially critical information are common side effects when changing grain size, particularly for categorical raster data sets (Gardner et al. 1982).

Two methods of spatial aggregation for categorical raster maps are standard with geographic information systems software packages: block majority filtering (MAJ) and nearest neighbor resampling (NN). Suppose an input map originally has a 30m x 30m grain size and the desired output map has a 90m x 90m grain size. MAJ aggregates by examining the category of the nine original pixels within each 90m x 90m aggregation unit and assigns the unit the category that occurs most often. NN is similar to a two-dimensional systematic sample in that every $k \times k$ unit of the raster map is sampled based on the center pixel. Following the example from above, a sample is taken on a three pixel by three pixel spacing to aggregate to a 90m x 90m pixel size. In addition to these methods, He et al. (2002) developed a random rule approach to spatial aggregation. This method works similar to the NN method except that the sample pixel randomly drawn from the pixels within the aggregation unit. Theoretically the NN and random rule methods should yield similar results however, the NN method does have potential bias. Because the NN method is a systematic sample, it can be biased if the classes exhibit a repetitive pattern with periodicity similar to the sample spacing (Steel et al. 1997). Beyond these issues, attention has been given to the general influence of spatially aggregating categorical raster maps on measures of amount and pattern (e.g., He et al. 2002, Moody and Woodcock 1995, Turner et al. 1989). In general, the MAJ aggregation technique forces dominant classes to become more dominant and rare classes to become rarer. However, the degree to which classes become more or less dominant depends on the spatial pattern of the original map. The NN method generally preserves the relative proportions but at the same time can degrade spatial pattern by introducing disaggregation (He et al. 2002). This often gives NN-aggregated maps a 'salt and pepper' appearance.

Categorical maps, regardless of the number of categories, can be thought of as a series of binary maps. Suppose a three-class map has developed, forest, and cultivated categories. Then, the three-class map is a combination of a developed / non-developed map, a forest / non-forest map, and a cultivated / non-cultivated map. These maps can be summarized by counts, both globally and within each aggregation unit. There are specific statistical models to deal with count data derived from a binary response; however, with few exceptions, these techniques have not been extended to the spatial domain. One technique developed for spatial and spatial-temporal count data is the spatial scan statistic (Kulldorff 1997). The spatial scan statistic was first developed for human epidemiology but has also been applied in ecology, brain imaging, psychology, toxicology, and veterinary medicine (Coulston and Riitters 2003, Yoshida 2003, Margai and Henry 2003, Sudakin et al. 2002, Hoar et al. 2003, respectively). The general objective of the scan statistic is to identify clusters of measurement units for which the occurrence of events is significantly more likely within the cluster than outside of the cluster. The scan statistic quantifies the importance of each potential cluster based on likelihood ratios and tests the significance of each potential cluster based on Monte Carlo simulation. Typically, the output clusters vary in size and shape. However, if the search for important potential clusters is omitted, and instead the map is divided into contiguous, non-overlapping squares of the same size then the resulting "clusters" may serve as equal area aggregation units. In turn, the scan statistic likelihood ratio can be calculated for each class within a given aggregation unit and this information can be used to determine the final classification of the unit when changing grain size.

The objectives of this study were to (1) implement the spatial scan statistic model to aggregate a 30m categorical map of land cover (2) describe the differences between aggregated maps created using the spatial scan statistic and other methods typically used for spatial aggregation and (3) suggest when the spatial scan statistic an appropriate method for spatial aggregation.

## 2 Materials and Methods

We used National Land Cover Data (NLCD) as the basis for our analysis (Vogelmann et al. 2001). The NLCD project used Landsat Thematic Mapper (TM) imagery (circa 1992) to classify the land cover of the coterminous United States in 21 classes at a grain size of 0.09 ha (30m by 30m). For the purposes of this study we collapsed the 21 original classes to seven broader classes (Table 1).

We selected a study area in southeastern Wisconsin, USA (43°17' N, 88°42' W) that encompassed approximately 13934 km2 (Figure 1). Based on the reclassified NLCD map, our study area was mostly cultivated (~ 72 %). Forest comprised approximately 12 percent of the study area, and wetland and urban categories each comprised six percent.

We used three methods of spatial aggregation—MAJ, NN, and the spatial scan statistic (SCAN)— to change the grain size of the original map from 0.09 ha pixels to 7.29 ha, 26.01 ha, and 98.01 ha pixels (representing 9×9, 17×17, and 33×33 blocks of the original pixels). SCAN can be applied using two different statistical models: the Bernoulli model and Poisson model. We used the Bernoulli model (Kulldorff 1997) for our analysis.

To implement this procedure the likelihood ratio was calculated for each land cover category within each aggregation unit.

The likelihood ratio based on the Bernoulli model was

$$\Psi_{az} = (c/n)^c \ ((n-c)/n)^{n-c} \ ((C-c)/(N-n))^{C-c} \ (((N-n)-(C-c))/(N-n))^{(N-n)-(C-c)} \ I \qquad [1]$$

where,

| Landcover (NLCD Classification) | Pixels | Percent | Connectivity |
|---|---|---|---|
| Water | 458528 | 2.96 | 0.89 |
| (11 Open Water) | | | |
| (12 Perennial Ice/Snow) | | | |
| Developed | 914122 | 5.90 | 0.80 |
| (21 Low Intensity Residential) | | | |
| (22 High Intensity Residential) | | | |
| (23 Commercial/Industrial/Transportation) | | | |
| Barren | 18711 | 0.12 | 0.81 |
| (31 Bare Rock/Sand/Clay) | | | |
| (32 Quarries/Strip Mines/Gravel Pits) | | | |
| (33 Transitional) | | | |
| Forested | 1853204 | 11.97 | 0.65 |
| (41 Deciduous Forest) | | | |
| (42 Evergreen Forest) | | | |
| (43 Mixed Forest) | | | |
| Grass/Shrubland | 152767 | 0.99 | 0.34 |
| (51 Shrubland) | | | |
| (71 Grasslands/Herbaceous) | | | |
| Cultivated | 11167042 | 72.13 | 0.94 |
| (61 Orchards/Vineyards/Other) | | | |
| (81 Pasture/Hay) | | | |
| (82 Row Crops) | | | |
| (83 Small Grains) | | | |
| (84 Fallow) | | | |
| (85 Urban/Recreational Grasses) | | | |
| Wetlands | 918001 | 5.93 | 0.70 |
| (91 Woody Wetlands) | | | |
| (92 Emergent Herbaceous Wetlands) | | | |
| Total | 15482375 | | |

**Table 1. Classification and characteristics of the study area at the original 0.09 ha grain size. Connectivity represents the probability that an adjacent pixel (4-neighbor rule) is of the same class.**

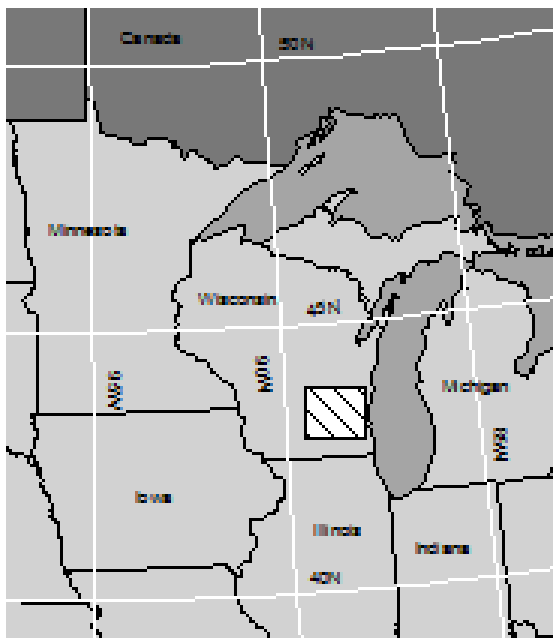

**Figure 1. The study area in southeastern Wisconsin, USA (represented by the diagonal hatch bars).**

$\Psi_{az}$ = the likelihood ratio for aggregation unit *a* and land cover category *z*
c = the number of pixels of category *z* within aggregation unit *a*
n = the number of pixels in each aggregation unit
C = the total number of pixels of category *z* within the study area
N = the total number of pixels in the study area.
I = indicator function where I = 1 if (c/n) ≥ (C/N) and zero otherwise.
The aggregation unit was then assigned the category with the greatest likelihood ratio value.

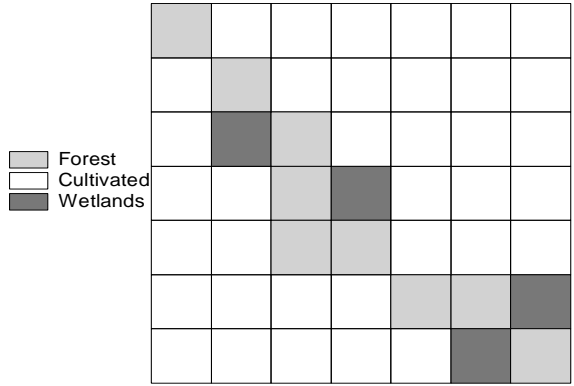

**Figure 2. Example of a 7×7 aggregation unit composed of cultivated land, forest, and wetlands.**

Use of the three aggregation techniques is best illustrated with a comparative example. Suppose Figure 2 represents a 7×7 pixel aggregation unit made up of forest, cultivated, and wetland cover types. Based on MAJ, this aggregation unit would be classified as cultivated because more pixels are classified as cultivated than either forest or wetlands. The NN method would instead classify this aggregation unit as wetlands because the center pixel (i.e., the location of the systematic sample point) is classified as wetlands. To apply the SCAN, we calculate the likelihood ratio for each category. The variables *c* and *n* are counts from the aggregation unit while the variables *C* and *N* are based on map-wide totals from Table 1. Notably, these variables allow the SCAN to incorporate both local and global (i.e., image-wide) information about each class during the aggregation process. For the forest category, *c*=9, *n*=49, *C*=1853204, and *N*=15482375. The likelihood ratio is generally computed in logarithmic form because of computer limitations on exponents, so $\log(\Psi_{forest})$ = 0.836. For the cultivated class, *c*=36, *n*=49, *C*=11167042, *N*=15482375, and $\log(\Psi_{cultivated})$=0.022. For the wetlands class, *c*=4, *n*=49, *C*=918001, *N*=15482375, and $\log(\Psi_{wetlands})$=0.197. Because the forest class has the greatest log likelihood ratio, the aggregation unit would be classified as forest by SCAN. To compare the three aggregation approaches, we investigated measures of both their global and local behavior. To describe the global behavior of each technique, we recorded the relative proportions of the map occupied by each land cover class at the three aggregation grain sizes. We examined local behavior using cumulative distribution functions (CDF) and agreement matrices. The CDFs allowed us to visually compare the local-scale characteristics of each method across land cover classes. To calculate the CDF for each aggregation technique and grain size, we computed the proportion ($P_z$) of original 0.9-ha pixels within each aggregation unit that were in the land cover category ultimately assigned to the unit by the aggregation method. Returning to the example aggregation unit in Figure 2, the $P_z$ for the MAJ technique—which classified the unit as cultivated—was 0.73. In contrast, the $P_z$ for the NN technique (where $z$ = wetlands) was 0.08 and the $P_z$ for the SCAN technique (where $z$ = forest) was 0.18. We then compiled the $P_z$ for each aggregation unit, aggregation method, and grain size was then used to create the CDFs.

We constructed agreement matrices to examine categorical overlap between maps of the same grain size on a pixel-by-pixel basis. Specifically, we examined overall agreement as well as Cohen's kappa ($K_{hat}$) across land cover classes for each grain size (Jensen 1996). We also examined individual land cover class agreement among aggregation methods.

## 3 Results

The MAJ, NN, and SCAN methods exhibited different trends when aggregating the original map to the 7.29, 26.01, and 98.01 ha grain sizes (Figure 3).

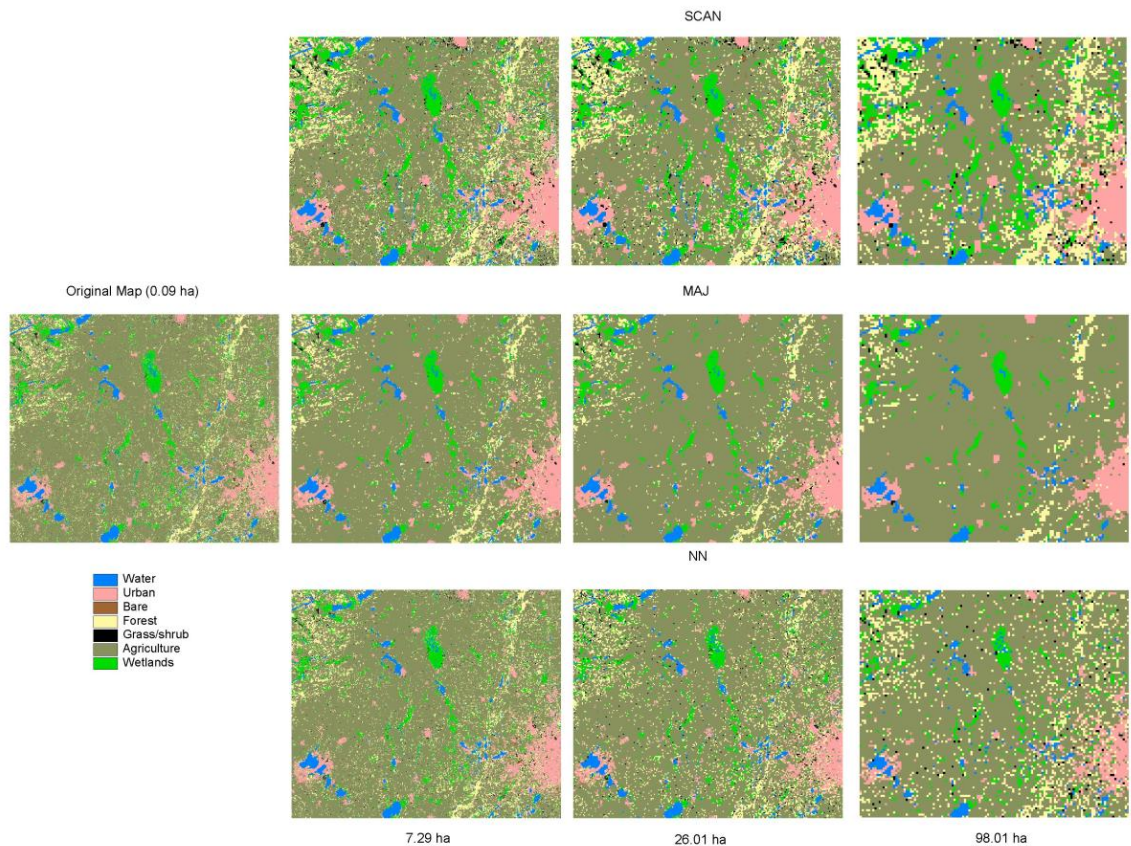

**Figure 3. The original seven-class land cover map and maps aggregated at the 7.29, 26.01, and 98.01 ha grain sizes using each aggregation technique.**

The MAJ method increased the proportion of the cultivated class (i.e., the most dominant class) by an average of 6.0 percent and decreased the proportions of rarer classes such as forest, water, and wetlands (Figure 4). The NN method tended to preserve the original proportions of each class across grain sizes, but the classes became less spatially cohesive. This was particularly evident at the 98.01 ha resolution, where a significant "salt-and-pepper" effect was observable, in contrast to the cohesiveness exhibited by the MAJ and SCAN techniques (Figure 3). The SCAN technique behaved in a manner inverse to the MAJ technique, emphasizing rare classes such as forest and wetlands (Figures 3 and 4). The proportions of these rarer classes increased at the expense of the more abundant classes such as cultivated, which represented ~72% of the landscape at the 0.09 ha grain size, but only ~52% of the landscape at the 98.01 ha resolution.

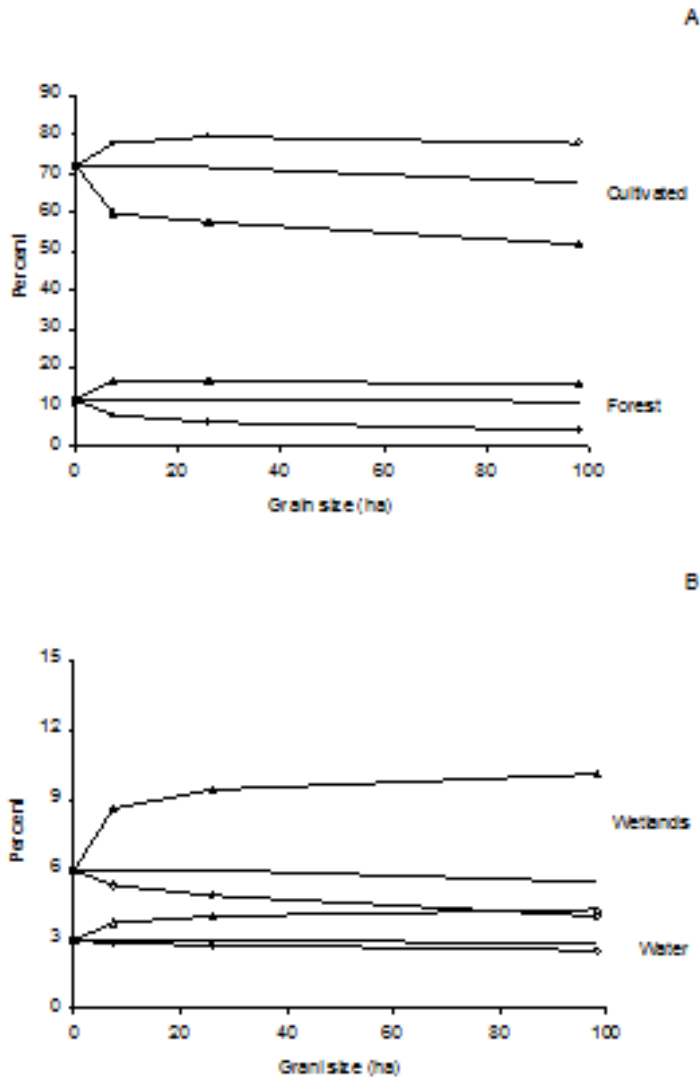

**Figure 4. Percent of the land cover map represented by (A) cultivated land and forest and (B) across grain sizes based on the MAJ method (open circle), the NN method (solid line), and the SCAN method (open triangle).**

Looking at the local behavior of each aggregation technique as represented by the CDFs (Figure 5), the techniques performed similarly when $P_z > 0.8$, regardless of grain size. In short, when any category $z$ represented more than 80% of an aggregation unit, the aggregation unit was classified as that category by all three techniques. The NN and SCAN techniques also performed similarly when $P_z < 0.2$. The largest difference between techniques occurred when $P_z \sim 0.45$: The disparity between the CDFs of the MAJ and the SCAN techniques grew as the grain size increased. Generally, when $P_z > 0.2$ and $P_z < 0.8$, the CDF for the SCAN technique was above the CDFs for the NN and MAJ techniques.

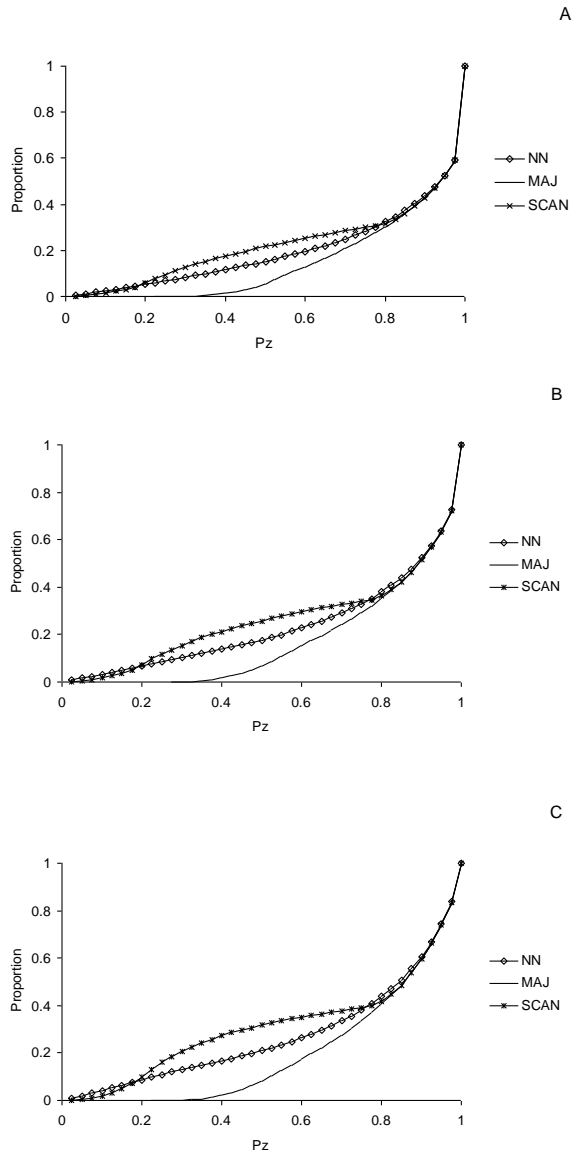

**Figure 5. Cumulative distribution functions (CDFs) for each aggregation method for the 7.29 ha grain size (A), the 26.01 ha grain size (B), and the 98.01 ha grain size (C).**

The SCAN method assigned a category to each aggregation unit (Equation 1) by comparing the proportion of category $z$ within each aggregation (i.e. $c_z/n_z$) to the overall proportion of category $z$ in the original map (i.e. $C_z/N_z$). Because of the indicator function $I$, an aggregation unit could only be classified as category $z$ if $c_z/n_z > C_z/N_z$ (the proportion of the category in the aggregation unit was greater than its overall proportion in the original map). When $c_z/n_z > C_z/N_z$ the log likelihood ratio was a measure of the concentration of the category within the aggregation unit compared to the concentration of the category in the entire map. This preserved rarer classes, which could have a high log likelihood ratio at relatively low values of $P_z$ (Figure 6). In contrast, for an aggregation unit to be classified as agriculture (the most abundant class) the proportion had to be greater than ~ 0.76 (Figure 6).

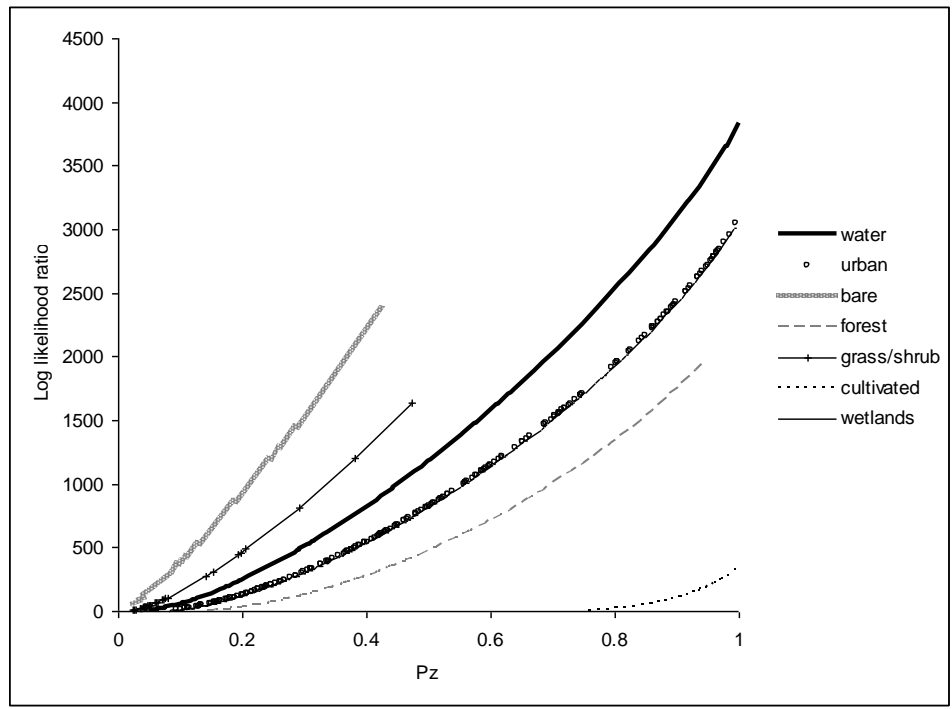

**Figure 6. Log likelihood ratio versus the proportion (Pz) of each aggregation unit, across grain sizes. The curves for the urban and wetland classes overlap because of the similar proportions occupied by each of these categories in the original map.**

Percent agreement between maps created using each aggregation method ranged from 81% agreement between the MAJ and SCAN methods at the 7.29 ha grain size to 71.2% agreement between the NN and SCAN methods at the 98.01 ha grain size (Table 2). In general, the agreement between the SCAN method and the MAJ and NN methods decreased as the grain size increased. Among all grain sizes, the overall percent agreement between the SCAN and the MAJ methods was higher than the agreement between the SCAN and NN methods, although the Khat coefficients suggested slightly higher agreement between the SCAN and the NN methods at the 26.01 ha and 98.01 ha grain sizes.

|              | MAJ       |           | NN        |           |
|--------------|-----------|-----------|-----------|-----------|
| Grain size   | Agreement | $K_{hat}$ | Agreement | $K_{hat}$ |
| ha           | %         | %         | %         | %         |
| 7.29         | 81.0      | 63.0      | 79.8      | 62.4      |
| 26.01        | 77.3      | 55.8      | 76.0      | 56.1      |
| 98.01        | 72.0      | 46.1      | 71.2      | 48.8      |

**Table 2.** Overall percent agreement and Khat comparing maps produced using the SCAN method with maps produced using the MAJ and NN methods across grain size.

|      |             | MAJ         |        |          |           |       |       |           | total  |        |
|------|-------------|-------------|--------|----------|-----------|-------|-------|-----------|--------|--------|
|      |             | grass/shrub | forest | wetlands | developed | bare  | water | cultivated| SCAN   |        |
|      | grass/shrub | 419         | 248    | 106      | 60        | 0     | 3     | 3667      | 4503   | 9.3%   |
|      | forest      | 0           | 13510  | 0        | 0         | 0     | 0     | 18203     | 31713  | 42.6%  |
|      | wetlands    | 0           | 613    | 9937     | 0         | 0     | 0     | 5906      | 16456  | 60.4%  |
| SCAN | developed   | 0           | 227    | 0        | 11302     | 0     | 0     | 5210      | 16739  | 67.5%  |
|      | bare        | 6           | 20     | 13       | 17        | 231   | 5     | 233       | 525    | 44.0%  |
|      | water       | 0           | 174    | 111      | 39        | 0     | 5310  | 1360      | 6994   | 75.9%  |
|      | cultivated  | 0           | 0      | 0        | 0         | 0     | 0     | 113618    | 113618 | 100.0% |
|      | total MAJ   | 425         | 14792  | 10167    | 11418     | 231   | 5318  | 148197    |        |        |
|      |             | 98.6%       | 91.3%  | 97.7%    | 99.0%     | 100.0%| 99.8% | 76.7%     |        | 81.0%  |

**Table 3.** Agreement matrix for the SCAN and MAJ aggregation techniques at the 7.29 ha grain size. Note that the land cover classes are in order of increasing connectivity (see Table 1).

|      |             | NN          |        |          |           |       |       |           | total  |        |
|------|-------------|-------------|--------|----------|-----------|-------|-------|-----------|--------|--------|
|      |             | grass/shrub | forest | wetlands | developed | bare  | water | cultivated| SCAN   |        |
|      | grass/shrub | 816         | 616    | 202      | 122       | 1     | 10    | 2736      | 4503   | 18.1%  |
|      | forest      | 442         | 15852  | 1140     | 228       | 0     | 91    | 13960     | 31713  | 50.0%  |
|      | wetlands    | 171         | 2393   | 9115     | 38        | 1     | 303   | 4435      | 16456  | 55.4%  |
| SCAN | developed   | 110         | 1177   | 163      | 10451     | 2     | 86    | 4750      | 16739  | 62.4%  |
|      | bare        | 14          | 54     | 13       | 19        | 224   | 7     | 194       | 525    | 42.7%  |
|      | water       | 28          | 433    | 399      | 146       | 0     | 5035  | 953       | 6994   | 72.0%  |
|      | cultivated  | 246         | 2258   | 366      | 176       | 3     | 28    | 110541    | 113618 | 97.3%  |
|      | total NN    | 1827        | 22783  | 11398    | 11180     | 231   | 5560  | 137569    |        |        |
|      |             | 44.7%       | 69.6%  | 80.0%    | 93.5%     | 97.0% | 90.6% | 80.4%     |        | 79.8%  |

**Table 4.** Agreement matrix for the SCAN and NN aggregation techniques at the 7.29 ha grain size. Note that the land cover classes are in order of increasing connectivity (see Table 1).

On an individual class basis, the SCAN method was more likely to assign pixels to the same category as the MAJ or NN methods when the category in the original map was more connected (Tables 3 and 4). For example the grass/shrub category had low connectivity in the original map (Table 1). At the 7.29 ha grain size, the SCAN method only had 9.3% agreement with the MAJ method (Table 3) and 18.1% agreement with the NN method (Table 4) for this class. In contrast, the cultivated category had high connectivity in the original map. Subsequently, the SCAN method had 100% agreement with the MAJ method and 97.3% agreement with the NN method at the 7.29 ha grain size. The barren class, however, did not follow the general trend of increased agreement with increased connectivity: It had a relatively high connectivity value, but the agreement between the SCAN method and the NN and MAJ methods was 44% and 42.7%, respectively. This was because the overall proportion of the barren class was 0.12% in the original map and therefore could exhibit a high log likelihood ratio even at low values of $P_z$ (Figure 6).

While all of the pixels classified as cultivated by the SCAN method were also classified as cultivated by the MAJ method, the MAJ method assigned a further ~34,500 pixels to the cultivated class at the 7.29 ha grain size (Table 3). Therefore, the MAJ method only agreed with the SCAN method 76.7% of the time for the cultivated class at the 7.29 ha scale. In general, agreement between the SCAN and MAJ methods decreased with increasing abundance of a given category in the original image.

**4 Discussion**

Changing scale by manipulating data can be a useful surrogate for observing the landscape directly with two or more sensors at different resolutions. However, the surrogate fundamentally differs from direct observation. Our understanding of the effects of "rescaling data" like in aggregation procedures is still rudimentary although some insight may be gained from a synthesis of numerous studies carried out in geography and remote sensing (e.g., Openshaw 1984; Justice et al. 1989; Jelinski and Wu 1996; Bian and Butler 1999; Wu 2004). More understanding can come out from the direct comparison of different aggregation techniques like those compared in this paper. Regardless of how data are changed subsequently at the scale of analysis, one must be cautious in interpreting results from rescaled data, and be aware that patterns and scales revealed in such analyses may not correspond to those in the real landscapes, or not even to those embodied into the data set the rescaling is based on.

Selecting an appropriate aggregation technique should be based on the goals of the particular study. No one technique is optimal for all situations, and each aggregation technique distorts different aspects of the original image. For instance, the MAJ technique is often used to approximate land cover information derived from a sensor with a coarser grain size, because it is seen as a reasonable simulation of how the coarser sensor behaves (He et al. 2002). Nonetheless, in studies relating to landscape heterogeneity, the SCAN and NN techniques may be more appropriate because they tend to preserve the presence of each class in some fashion, while the MAJ technique tends to lose rare classes completely (Turner et al. 1989). Ecologists are often interested in relating the broad-scale patterns of sparsely or sporadically distributed habitat within a matrix of mostly non-habitat to bioclimatic, topographic, and edaphic factors derived from information of a different grain (e.g., Coulston and Riitters 2005). In such cases, the SCAN technique may be most appropriate because it maintains and enhances the presence of less abundant land cover classes during aggregation.

Commonly accepted methods of spatially aggregating categorical raster maps do not use both local and global information. The majority rule uses only local information and is biased towards more abundant and highly connected classes (Turner et al. 1989). Nearest neighbor approaches, as well as the random rule approach suggested by He et al. (2002), preserve relative proportions at a global scale because they behave similarly to a simple random sample. However, they tend to produce spatially disaggregated patterns (He et al. 2002). Turner et al. (1989) suggested that developing methods that preserve information across spatial scales is critical. The spatial scan statistic can be used to spatially aggregate classified satellite imagery and preserve information from less abundant classes by incorporating both global and local scale information.

There are several landscape metrics that describe the spatial pattern of categorical raster maps. Based on a multivariate factor analysis, Riitters et al. (1995) found that commonly used landscape metrics represent approximately six independent dimensions of pattern in categorical raster maps. These six dimensions can be represented by the following univariate metrics: average perimeter-area ratio, contagion, standardized patch shape, patch perimeter-area scaling, number of attribute classes, and large-patch density-area scaling. With the exception of the number of attribute classes, comparing these metrics across grain size is problematic and may be invalid because the results reflect scale-related errors rather than true differences in pattern (Turner et al. 2001). For example, aggregating maps, regardless of technique, creates larger patches of a more uniform shape unless the class ceases to exist, and patch-based metrics will be artificially changed as a result (Turner et al. 1989, Turner et al. 2001, He et al. 2002). Contagion is also difficult to interpret across grain sizes; in particular, the direction of change in contagion metrics due to spatial aggregation depends on whether area is taken into account. For example suppose an input map has a grain size of 0.09 ha and we change the grain size of the map to 7.29 ha. If we define contagion as the probability that a pixel of one class is next to a pixel of the same class, then contagion decreases with spatial as grain size increases. However, if contagion is considered the probability that a 0.09 ha square block of land of one class is next to a 0.09 ha square block of land of the same class then contagion increases with spatial aggregation because each 7.29 ha aggregation unit is comprised of 81 0.09 ha square blocks. Results not shown here demonstrated that the direction of change in patch-based and contagion-based metrics with increasing grain size was similar for the three aggregation techniques, although the magnitude was slightly different among aggregation techniques.

In this paper we introduced the spatial scan statistic as a technique to aggregate categorical raster maps to coarser resolutions using fixed windows. The spatial scan statistic can also be used in a sliding window capacity for digital image post-processing. It is a generally accepted practice to use a filter to reduce speckle in an image classified on a per-pixel basis (Davis and Peet 1997, Goodchild 1994, ERDAS 2003). In cases where these "speckles" represent classes of particular interest, the spatial scan statistic can be used as a filter to enhance rare classes. For example, Koch (2005) created a decision tree classifier for mapping eastern hemlock (*Tsuga canadensis*) stands in the southern Appalachian Mountains based on Advanced Spaceborne Thermal Emission and Reflection Radiometer imagery and ancillary data. The resulting stand map was a key piece of information for identifying stands at risk of infestation by an insect pest, hemlock woolly adelgid (*Adelges tsugae*), and prioritizing them for control efforts. While hemlocks are scattered throughout the region, distinct stands are relatively rare features in the southern Appalachians, representing ~2% of the U.S. total (McWilliams and Schmidt 1999). Using the spatial scan statistic as a filter to enhance the pattern of hemlock would reduce omission errors as well as yield a map emphasizing the proximity and potential connectivity of hemlock stands for enabling adelgid spread.

There are several techniques available to spatially aggregate categorical raster maps. The spatial scan statistic is a technique which can be used with count data derived from a binary response such as categorical raster maps. This technique uses information from both global and local scales and compares the relative importance among land cover classes to spatially aggregate raster maps. In situations where it is necessary to preserve information from less abundant classes the spatial scan statistic provides a viable alternative to nearest neighbor resampling and block majority filtering.

**Acknowledgments**

The research described in this article was supported in part by a Cooperative Agreement between North Carolina State University and the U.S. Forest Service.  Computing facilities were provided by the Center for Landscape Pattern Analysis.

**References**

Bian L. and Butler R. 1999. Comparing effects of aggregation methods on statistical and spatial properties of simulated spatial data. Photogrammetric Engineering and Remote Sensing 65: 73–84.

Coulston, J.W., Riitters, K.H.  2005.  Preserving biodiversity under current and future climates: a case study.  Global Ecology and Biogeography 14: 31-38.

Coulston, J.W., Riitters, K.H. 2003. Geographic analysis of forest health indicators using spatial scan statistics. Environmental Management 31: 764-773.

Davis, W.A., Peet, F.G. 1977. A method of smoothing digital thematic maps. Remote Sensing of Environment 6: 45-49.

ERDAS. 2003.  ERDAS Field Guide, seventh edition.  Leica Geosystems GIS & Mapping, LLC, Atlanta GA.  672 pp.

Gardner, R.H., Cale, W.G., O'Neill, R.V.  1982.  Robust analysis of aggregation error.  Ecology 63:  1771-1779.

Goodchild, M.F.  1994.  Integrating GIS and remote-sensing for vegetation and analysis and modeling- methodological issues.  Journal of Vegetation Science 5: 615-626.

He, H.S., Ventura, S.J., Mladenoff, D.J.  2002.  Effects of spatial aggregation approaches on classified satellite imagery.  International Journal of Geographic Information Science 16: 93-109.

Hoar, B.R., Chomel, B.B., Rolfe, D.L., Chang, C.C., Fritz, C.L., Sacks, B.N., Carpenter, T.E. 2003.  Spatial analysis of *Yersinia pestis* and *Bartonella vinsonii* subsp. *berkhoffii* seroprevalence in California coyotes (*Canis latrans*). Preventive Veterinary Medicine 56: 299-311.

Jelinski D.E. and Wu J. 1996. The modifiable areal unit problem and implications for landscape ecology. Landscape Ecology 11: 129–140.

Jensen, J.R. 1996.  Introductory Digital Image Processing: A Remote Sensing Perspective.  Prentice Hall, Upper Saddle River, NJ.  318 pp.

Justice C.O., Markham B.L., Townshend J.R.G. and Kennard R.L. 1989. Spatial degradation of satellite data. International Journal of Remote Sensing 10: 1539–1561.

Koch, F.H.  2005.  Spatial Tools for Managing Hemlock Woolly Adelgid in the Southern Appalachians.  PhD Dissertation.  North Carolina State University, Raleigh, NC.  225 pp.

Kulldorff, M. 1997. A spatial scan statistic. Communications in Statistics: Theory and Methods 26:1481–1496.

Li, H., Reynolds, J.F.  1995.  On definition and quantification of heterogeneity.  Oikos 73: 280-284.

Margai, F., Henry, N. 2003.  A community-based assessment of learning disabilities using environmental and contextual risk factors. Social Science and Medicine 56: 1073-1085.

McWilliams, W.H., Schmidt, T.L. 1999.  Composition, structure, and sustainability of hemlock ecosystems in eastern North America.  Pages 5-10 in: K. A. McManus, K.S. Shields, D. R. Souto, eds., Proceedings: Symposium on Sustainable Management of Hemlock Ecosystems in Eastern North America, June 22-24, 1999, Durham, NH.  USDA Forest Service, Northeastern Research Station, General Technical Report NE-267.

Moody, A., Woodcock, C.E.  1995.  The influence of scale and the spatial characteristics of landscapes on land-cover mapping using remote sensing.  Landscape Ecology 6:  363-379.

Openshaw S. 1984. The Modifiable Areal Unit Problem. GeoBooks, Norwich, UK. of Biology 70: 439–466.

Riitters, K.H., O'Neill, R.V., Hunsaker, C.T., Wickham, J.D., Yankee, D.H., Timmins, S.P., Jones, K.B., Jackson, B.L.  1995.  A factor analysis of landscape pattern and structure metrics.  Landscape Ecology 10:  23-39.

Steel, R.G.D, Torrie, J.J., Dickey, D.A.  1997.  Principles and Procedures of Statistics:  A Biometrical Approach, Third Edition.  The McGraw-Hill Companies, Inc.  New York. 666 pp.

Sudakin, D.L., Horowitz, Z., Giffin, S. 2002.  Regional variation in the incidence of symptomatic pesticide exposures: Applications of geographic information systems.  Journal of Toxicology - Clinical Toxicology 40: 767-773.

Turner, M.G., O'Neill, R.V., Gardner, R.H., Milne B.T.  1989.  Effects of changing spatial scale on the analysis of landscape pattern.  Landscape Ecology 3: 153-162.

Turner, M.G., Gardner, R.H., O'Neill, R.V.  2001.  Landscape Ecology In Theory and Practice – Pattern and Process. Springer-Verlag, New York.  401 pp.

Verburg, P.H., Veldkamp, A.  2004.  Projecting land use transitions at forest fringes in the Philippines at two spatial scales.  Landscape Ecology 19: 77-98.

Vogelmann, J. E., Howard, S.M., Yang, L., Larson, C.R., Wylie, B.K., Driel, N. 2001. Completion of the 1990s national land cover data set for the conterminous United States from Landsat Thematic Mapper data and ancillary data sources. Photogrammetric Engineering & Remote Sensing 67: 650–662.

Wu J. 2004. Effects of changing scale on landscape pattern analysis: Scaling relations. Landscape Ecology 19: 125–138.

Yoshida, M., Naya, Y., Miyashita, Y. 2003.  Anatomical organization of forward fiber projections from area TE to perirhinal neurons representing visual long-term memory in monkeys. Proceedings of the National Academy of Sciences of the United States of America 100: 4257-4262.